\documentclass[aps,prl,reprint,nobalancelastpage,nofootinbib]{revtex4-2}
\pdfoutput=1

\usepackage[utf8]{inputenc}

\usepackage{amsmath}
\usepackage{amssymb}
\usepackage[all]{xy}
\usepackage{tikz}
\usetikzlibrary{calc}
\usepackage{pgfplots}
\pgfplotsset{/pgf/number format/use comma, compat=newest}
\usepackage{xcolor}
\usepgfplotslibrary{fillbetween}

\usepackage[toc,page]{appendix}
\usepackage{enumerate}
\usepackage{tensor}
\usepackage{booktabs}
\usepackage{bbm}
\usepackage{youngtab}
\usepackage{slashed}
\usepackage{multirow}
\usepackage{makecell}
\usepackage{float}
\usepackage{comment}
\usepackage{verbatim}
\usepackage{xcolor}
\usepackage[framemethod=tikz]{mdframed}
\usepackage{soul}
\usepackage{mathtools}
\usepackage[inline]{enumitem}
\usepackage{alphabeta}
\usepackage{accents}
\usepackage{soul}

\usepackage{amsthm}

\newcommand{\ab}{|}
\newcommand{\der}{\partial}
\newcommand{\de}{\mathrm{d}}

\newcommand{\s}{
{\mathrm{sp}}}
\newcommand{\DM}{
{\mathrm{DM}}}

\newcommand{\tdelta}{\tilde{\delta}}
\newcommand{\tsigma}{\tilde{\sigma}}

\newcommand{\e}{\mathrm{e}}

\newcommand{\p}{\mathrm{P}}

\allowdisplaybreaks

\usepackage{hyperref}
\hypersetup{
    colorlinks=true,
    linkcolor=purple,
    citecolor=purple,
    filecolor=purple,      
    urlcolor=purple,
}

\makeatletter
\renewcommand{\p@subsection}{}
\renewcommand{\p@subsubsection}{}
\makeatother

\begin{document}

\title{Long-lived SEC violation via DM/DE couplings}

\author{Gary Shiu}
\email{shiu@physics.wisc.edu}
\affiliation{Department of Physics, University of Wisconsin-Madison, 1150 University Avenue, Madison, WI 53706, USA}

\author{Flavio Tonioni}
\email{flavio.tonioni@unipd.it}
\affiliation{Dipartimento di Fisica e Astronomia, Università degli Studi di Padova, and INFN-Padova, via F. Marzolo 8, 35131 Padua, Italy}

\author{Hung V. Tran}
\email{hung@math.wisc.edu}
\affiliation{Department of Mathematics, University of Wisconsin-Madison, 480 Lincoln Drive, Madison, WI 53706, USA}

\begin{abstract}
We discuss a cosmological scenario where an effective violation of the strong energy condition (SEC) is realized through a coupling between SEC-fulfilling dark matter (DM) and dark energy (DE).
Although the SEC-violating solutions might in principle last for an arbitrarily long time, we highlight several challenges that string realizations must face: most notably, these are the identification of suitable heavy states and their relationship with the theory cutoff.
Furthermore, we discuss a black-hole argument that still allows for long-lived epochs of cosmic acceleration,  but that prevents them from lasting forever.
We also discuss negative potentials in the presence of a tower of light states, showing that the DM/DE coupling can push the theory towards regions of parametric control.
\end{abstract}

\maketitle

\section{Introduction}

The nature of dark energy (DE) is arguably one of the deepest puzzles in fundamental physics, but a fully explicit realization in a ultraviolet-complete (UV-complete) theory of gravity remains elusive \cite{Danielsson:2018ztv}.
Whether DE is time-varying or constant,
the criterion for cosmic acceleration can be universally stated in terms of the Hubble parameter $\smash{H = \dot{a}/a}$, where $a$ is the cosmic scale factor and derivatives are with respect to cosmological time.
The universe accelerates, i.e. $\smash{\ddot{a} >0}$, if and only if $\smash{\epsilon = - \dot{H}/H^2 <1}$, which is equivalent to the strong energy condition (SEC) being violated.

Various energy conditions have been proposed to exclude potentially unphysical spacetimes.
For example, the null energy condition (NEC) is sufficient to prove the absence of traversable wormholes \cite{Morris:1988cz}.
It is also crucial in the Hawking-Penrose singularity theorem \cite{Hawking:1970zqf} and in the second law of black-hole thermodynamics \cite{Bardeen:1973gs}.
Unlike the NEC, whose violation generally leads to pathologies like loss of unitarity, SEC violation seems relatively harmless.
For instance, inflationary cosmology violates the SEC.
Yet, the scalar potentials in UV-complete theories such as those arising in string compactifications are arguably too steep to accommodate extended phases of cosmic acceleration, at least for spatially-flat cosmologies in asymptotic regimes of the field space \cite{Obied:2018sgi, Ooguri:2018wrx, Hebecker:2018vxz}.
One can quantify the steepness of the potential $V$ in a $d$-dimensional theory via the criterion $\ab \der V \ab / \kappa_d V \geq c_V$, where $\der V$ is the potential gradient and $c_V$ is an order-1 constant \cite{Obied:2018sgi, Ooguri:2018wrx}, with $\smash{\kappa_d = 1/m_{\p, d}^{d/2-1}}$ the gravitational coupling.
For fields evolving towards the field-space asymptotics, $\smash{c_V = 2 / \sqrt{d-2}}$ is the proposed value of the constant \cite{Bedroya:2019snp, Rudelius:2021azq, Shiu:2023nph}, which is incompatible with arbitrarily long-lived cosmic acceleration.
This is because in the limiting case, where the inequality is saturated, one has $\smash{\epsilon = [(d-2)/4] \, c_V^2}$.
In this note, we discuss a mechanism to achieve a long-lived epoch with $\epsilon<1$, despite an overly-steep potential.
In spatially-hyperbolic cosmologies, arbitrarily long-lived epochs of cosmic acceleration may exist at the field-space boundary \cite{Marconnet:2022fmx, Andriot:2023wvg, Marconnet:2025vhj}.
These would always approach the value $\epsilon=1$, and would still not feature a cosmic horizon.
The solutions we study in principle admit arbitrarily small values of $\epsilon$, and a cosmic horizon, at the cost, however, of generating black holes whose horizon would outgrow the observable universe.
We will discuss this point in detail below.

Our idea is as follows. A coupling of dark matter (DM) to DE may induce transient effective NEC violations on the DE fluid \cite{Das:2005yj}.\footnote{We stress that the NEC violation owes to the arbitrariness of considering the interaction term as part of the DE or DM energy-momentum tensor.
The total energy-momentum tensor fulfills the NEC at all times, but a convenient parameterization of the total energy results in a DE term violating the NEC.}
This DM/DE coupling has recently been revisited \cite{Chakraborty:2025syu, Khoury:2025txd, Andriot:2025los} in light of the DESI results \cite{DESI:2025fxa, DESI:2025zpo, DESI:2025zgx}.
The energy density of a cosmological fluid with constant state parameter $w$ evolves as $\rho = \rho_0 \, (a_0/a)^{(d-1) (1+w)}$, where subscripts denote initial conditions.
A NEC-violating fluid thus has a growing energy density.
Indeed, if there is a second fluid decaying into DE, then effectively the DE density may increase over time while still fulfilling the NEC.
Our observation is that, just like the DM/DE coupling may induce an effective DE $w$-parameter below the NEC threshold, it can also induce a SEC violation -- in this case for both the DE and DM fluids --, even though each component of the universe would by itself fulfill the SEC, and it can do so for an extended period of time.
Note that the mechanism we propose differs from transient NEC violations in two respects:
\begin{enumerate*}[label=(\roman*)]
    \item we are aiming for a prolonged duration of acceleration;
    \item it is not our goal to realize phantom DE, with $w<-1$.
\end{enumerate*}

In this note, we discuss cosmological solutions featuring this DM/DE coupling, which in fact date back to refs.~\cite{Wetterich:1994bg, Amendola:1999qq, Amendola:1999er, Billyard:2000bh}.
Recently, phenomenological implications of generalized versions of such DM/DE-coupled models were explored by refs.~\cite{Amendola:2020ldb, Liu:2023mwx}.
We also comment on similar but different scenarios to realize cosmic acceleration with DM/DE couplings explored in refs.~\cite{Khoury:2003aq, Khoury:2003rn, Brax:2004px}, ref.~\cite{Agrawal:2019dlm}, and refs.~\cite{Gomes:2023dat, Casas:2024xqy, Casas:2024oak}.
Then, we discuss challenges in realizing the mechanism in UV-complete models, suggesting promising ways out.
Even in the absence of fine-tuned initial conditions, the SEC-violating solutions can in principle be long-lived, i.e. lasting for a number of Hubble times.
However, a black-hole argument that we illustrate actually prevents these from being semi-eternal.
Eventually, we lay out the cosmological solutions for multi-dimensional DE field spaces, beyond the single-field results \cite{Wetterich:1994bg, Amendola:1999qq, Amendola:1999er, Billyard:2000bh}.

A compelling spin-off of our current study is an alternative mechanism realized via the DM/DE coupling.
We consider light towers of states in the presence of negative potentials.
While such potentials would drive the theory into the field-space interior, losing parametric control, we show that a coupling to light towers of states can instead push the fields towards the boundary.

\section{DM/DE-coupled cosmology}

DM may be composed of a gas of non-relativistic particles of mass $m$.
If so, the DM energy density scales inversely with the volume, via the number density $n_\DM$, and proportionally to the mass.
Let us assume the mass is not a constant, but rather a function $m(\varphi)$ of a field $\varphi$.
Then, to accommodate for the solution \cite{Das:2005yj}
\begin{equation}
    \rho_\DM = n_{\DM, 0} \, \Bigl(\dfrac{a_0}{a}\Bigr)^{d-1} m(\varphi),
\end{equation}
the continuity equation must be $\dot{\rho}_\DM + (d-1) H \rho_\DM = \rho_\DM \, \dot{\varphi} \, m'/m$, with $' = \der/\der \varphi$.
Consistency with the usual Friedmann equation requires the scalar field equation to also change, i.e. $\smash{\ddot{\varphi} + (d-1) H \dot{\varphi} + V' = - \rho_\DM \, m'/m}$.
Hence, practically, the effect of the coupling on the field is the generation of the effective potential
\begin{equation} \label{effective DMDE-coupled potential}
    U(\varphi) = V(\varphi) + n_{\DM, 0} \, \Bigl(\dfrac{a_0}{a}\Bigr)^{d-1} m(\varphi).
\end{equation}
If $V$ and $m$ have opposite slopes, the field experiences an effective potential minimum; see fig. \ref{fig: effective potential}.
Due to the dependence on the scale factor, such a minimum shifts with time.
Yet, the kinetic energy of the field, and hence the rate of deceleration, can be significantly reduced compared to the one that would be induced by $V$ alone.
This motivates us to consider $\varphi$ as the field realizing DE.

\begin{figure}[ht]
    \centering
    
    \begin{tikzpicture}[xscale=0.95,yscale=1.25]

    \draw[->] (-0.5,0) -- (6.5,0) node[below]{$\phi$};
            \draw[->] (0,-0.35) -- (0,2.65) node[left]{};

            \draw[domain=0.28:6.2, smooth, thick, variable=\x, magenta] plot ({\x}, {2.65*exp(-\x)}) node[above, black]{$V$};
    
            \draw[domain=0.28:5.0, smooth, thick, variable=\x, orange!80!purple] plot ({\x}, {2.65*exp(-\x) + 0.18*exp(0.5*\x)}) node[above, black]{$U(a_1)$};

            \draw[->,thin,gray!50!white] (3.6,1.0) -- (4.45,0.50) node[right,pos=0.2,black]{\;$a_2 > a_1$};
    
            \draw[domain=0.28:5.8, smooth, thick, variable=\x, orange!60!yellow] plot ({\x}, {2.65*exp(-\x) + 0.04*exp(0.5*\x)}) node[right, black]{$U(a_2)$};

    \end{tikzpicture}
    
    \caption{The potential $V = \Lambda \, \e^{- \kappa_d \gamma \varphi}$ (magenta) and the effective potential $\smash{U = V + n_{\DM, 0} \, (a_0/a)^{d-1} \, \mu \, \e^{\kappa_d \beta \varphi}}$ for a smaller (darker orange) and larger scale factor (lighter orange).}
    
    \label{fig: effective potential}
\end{figure}

Now, let us consider a universe with two components:
\begin{enumerate*}[label=(\roman*)]
    \item a barotropic fluid, with energy density $\rho$ and state parameter $w \in\; ]-1,1[$, which, for generality, we do not fix for now;
    \item a scalar sector, with a field $\varphi$ and a potential $V$.
\end{enumerate*}
In the presence of the fluid/field coupling, the cosmological equations read
\begin{subequations}
\begin{align}
    & \ddot{\varphi} + (d-1) H \dot{\varphi} + V' = - \rho \, \dfrac{m'}{m}, \label{DMDE FRW-KG eq.} \\
    & \dot{\rho} + (d-1) (1+w) H \rho = \rho \, \dot{\varphi} \, \dfrac{m'}{m}, \label{DMDE continuity eq.} \\
    & H^2 = \dfrac{2 \kappa_d^2}{(d-1) (d-2)} \biggl[ \dfrac{1}{2} \, \dot{\varphi}^2 + V + \rho \biggr], \label{DMDE Friedmann eq.}
\end{align}
\end{subequations}
also implying $\smash{\dot{H} = - [2 \kappa_d^2/(d-2)] \, \bigl[ \dot{\varphi}^2/2 + [(1+w)/2] \, \rho \bigr]}$.
From now on, we focus on exponential couplings
\begin{subequations}
    \begin{align}
        m & = \mu \, \e^{\kappa_d \beta \varphi}, \label{m} \\
        V & = \Lambda \, \e^{- \kappa_d \gamma \varphi}, \label{V}
    \end{align}
\end{subequations}
where $\beta, \gamma$ are constants and $\mu, \Lambda > 0$ are fixed reference values.
Without loss of generality, we fix $\gamma > 0$.
In string compactifications, exponential behaviors are in fact ubiquitous, through string-coupling and volume expansions.

We define $h = (d-1) H$, $b = \beta/\Gamma_d$ and $c = \gamma/\Gamma_d$, with $\smash{\Gamma_d = 2 \sqrt{d-1}/\sqrt{d-2}}$, and
\begin{align*}
    x & = \dfrac{\kappa_d}{\sqrt{d-1} \sqrt{d-2}} \, \dfrac{\dot{\varphi}}{H}, \\
    y & = \dfrac{\kappa_d \sqrt{2}}{\sqrt{d-1} \sqrt{d-2}} \, \dfrac{1}{H} \, \sqrt{\Lambda \, \e^{- \kappa_d \gamma \varphi}}, \\
    z & = \dfrac{\kappa_d \sqrt{2}}{\sqrt{d-1} \sqrt{d-2}} \, \dfrac{\sqrt{\rho}}{H},
\end{align*}
Then, eqs.~(\ref{DMDE FRW-KG eq.}, \ref{DMDE continuity eq.}, \ref{DMDE Friedmann eq.}) become \cite{Copeland:1997et, Amendola:1999qq, Amendola:1999er}
\begin{subequations}
\begin{align}
    \dot{x} & = \biggl[ - (x - c) (y)^2 - \dfrac{1-w}{2} \, x (z)^2 - b (z)^2 \biggr] \, h, \label{DMDE x-equation} \\
    \dot{y} & = \biggl[ (x)^2 + \dfrac{1+w}{2} \, (z)^2 - c x \biggr] \, y h, \label{DMDE y-equation} \\[0.45ex]
    \dot{z} & = \biggl[ - \dfrac{1+w}{2} + (x)^2 + \dfrac{1+w}{2} \, (z)^2 + b x \biggr] \, z h, \label{DMDE z-equation} \\
    \dot{h} & = - \biggl[ (x)^2 + \dfrac{1+w}{2} \, (z)^2 \biggr] h^2, \label{DMDE h-equation}
\end{align}
\end{subequations}
jointly with the condition
\begin{equation}
    (x)^2 + (y)^2 + (z)^2 = 1. \label{DMDE sphere condition}
\end{equation}
One can search for linearly-stable solutions by linearizing the equations around the critical points $(\mathrm{x}, \mathrm{y}, \mathrm{z})$ such that $\smash{(\dot{x}, \dot{y}, \dot{z}) = (0,0,0)}$ \cite{Copeland:1997et}.
Besides those where $\mathrm{y}=0$ and/or $\mathrm{z}=0$, the equations admit the critical point
\begin{subequations}
    \begin{align}
        \mathrm{x} & = \dfrac{1}{2} \, \dfrac{1+w}{b + c}, \label{xCP} \\[1.15ex]
        (\mathrm{y})^2 & = \dfrac{4b(b+c) + (1-w) (1+w)}{4 (b+c)^2}, \label{yCP} \\
        (\mathrm{z})^2 & = \dfrac{2c(b+c) - (1+w)}{2 (b+c)^2}, \label{zCP}
    \end{align}
\end{subequations}
with the existence conditions $c (b+c)>(1+w)/2$ and $b (b+c) + (1-w^2)/4>0$.
This critical point is linearly stable whenever it exists \cite{Amendola:1999qq, Amendola:1999er}.
In physical variables, it corresponds to
the time-dependent DM/DE-coupled solution
\begin{subequations}
    \begin{align}
        \varphi(t) & = \varphi_0 + \dfrac{2}{\kappa_d \gamma} \, \ln \dfrac{t}{t_0}, \label{DMDE phi} \\
        \rho(t) & = \dfrac{4}{(1+w) \gamma^2} \dfrac{1}{\kappa_d^2 t^2} \, \biggl[ \dfrac{1}{2} \dfrac{d-2}{d-1} \dfrac{\gamma (\gamma + \beta)}{1+w} - 1 \biggr], \label{DMDE rho}
    \end{align}
\end{subequations}
for suitable constants $\varphi_0$ and $t_0$, with
\begin{equation} \label{DMDE epsilon}
    \epsilon = \dfrac{d-1}{2} \dfrac{1+w}{1 + \beta/\gamma}.
\end{equation}
If $\beta>0$, for a sufficiently large ratio $\beta/\gamma$, the $\epsilon$-parameter can be arbitrarily small.
Additional solutions exist; for details, see app. \ref{app.: linear stability}.
Below, we only focus on the solution in eqs.~(\ref{DMDE phi}, \ref{DMDE rho}, \ref{DMDE epsilon}) because it is the only one that admits $\epsilon<1$ while fulfilling the inequality $\gamma^2 > 4/(d-2)$.

We note that the solutions in eqs.~(\ref{DMDE phi}, \ref{DMDE rho}, \ref{DMDE epsilon}) are different from those in refs.~\cite{Khoury:2003aq, Khoury:2003rn, Brax:2004px, Das:2005yj}, referred to inverse-power law potentials, while ref.~\cite{Agrawal:2019dlm} studies $\beta<0$.
Also, refs.~\cite{Gomes:2023dat, Casas:2024xqy} discuss transient epochs of cosmic acceleration, computed as quasi-de Sitter (dS) solutions through expansions near the minima of a DM/DE-coupled effective potential, while the solutions in eqs.~(\ref{DMDE phi}, \ref{DMDE rho}) are exact and linearly-stable.
In particular, the latter require no fine tuning of the initial conditions and are not transient.
Finally, as discussed below, by construction they do not need an additional field within the effective field theory (EFT) to realize DM.

\section{Seeking string embeddings}

To motivate the string-theoretic perturbative expansions, one generally needs large field values.
In eq.~(\ref{DMDE phi}), the field evolves with $\dot{\varphi}>0$, and, if $\beta > 0$, DM gets heavier and heavier.
In string compactifications, the EFT cutoff must lie below the species scale \cite{Veneziano:2001ah,Dvali:2007hz}, which is expected to fall as $\Lambda_\s = m_{\p, d} \, \e^{- \kappa_d \lambda \varphi}$, with $\lambda$ an order-1 constant \cite{Calderon-Infante:2023ler,vandeHeisteeg:2023ubh,vandeHeisteeg:2023uxj}.
Hence, at best, the DM mass hits the cutoff within a finite field-space distance, i.e. after a finite number of Hubble times.

However, we highlight the following: since we are treating DM as a cosmological fluid made out of non-relativistic matter constituents, it is not necessary for the DM mass to be below the cutoff.
One does not need a microscopic field to describe DM: the DM fluid can be treated as a classical source of energy density.
For instance, DM may consist of dark macroscopic objects that are bound configurations of dark-sector particles.
If such particles have a field-dependent mass, DM is expected to have one, as well.
Hence, although we cannot calculate the exact functional form, lacking of details on e.g. binding energy, we can still parametrize the DM mass as $m(\varphi)$.
In our analysis, the assumed exponential behavior allows us to handle a stable analytic solution, but we expect the phenomenon to be general.
In addition, as $m/H = (m_0/H_0) \, (t/t_0)^{2 \beta/\gamma + 1}$, the DM Compton wavelength falls down quickly with respect to the horizon size.
This makes any quantum effects dynamically less and less relevant with respect to our classical approximation.
Finally, as obvious from eq.~(\ref{DMDE rho}), although the DM mass grows exponentially, the universe expansion is still sufficient to dilute the DM energy density over time.
To add another perspective, we note that, although the DM particle eventually becomes heavier than the cutoff, the fact that it helps cosmic acceleration may be interpreted as a ``renormalization'' effect that manifests in an effective flattening of the potential.

Of course, the DE degrees of freedom must instead always lie within the EFT cutoff.
Both the kinetic and the potential energy fall over time with the same parametric dependence as the squared Hubble constant \cite{Shiu:2023fhb}.
Hence, we may simply require that $\smash{\kappa_d \sqrt{V} \leq \Lambda_\s}$ via eq.~(\ref{DMDE Friedmann eq.}) \cite{Hebecker:2018vxz}.

In the asymptotics of string compactifications, the potential slope is expected to be bounded as $\gamma \geq 2/\sqrt{d-2}$ \cite{Bedroya:2019snp, Rudelius:2021azq, Shiu:2023nph}.
Below, we seek realizations of suitable growing masses, too, setting $w=0$.
According to the distance conjecture \cite{Ooguri:2006in}, string compactifications involve towers of states that become light as one approaches the field-space asymptotics $\varphi = \infty$, with a mass gap
\begin{equation}
    m_{\mathrm{DC}}(\varphi) = \mu_{\mathrm{DC}} \, \e^{- \kappa_d \alpha \varphi},
\end{equation}
for an order-1 constant $\alpha > 0$ and a reference mass $\mu_{\mathrm{DC}}$.
Arguments exist supporting the bound $\alpha \geq 1/\sqrt{d-2}$ for the gap rate \cite{Etheredge:2022opl}.
Due to string dualities, towers of heavy states emerging in the opposite regime, i.e. $\varphi=-\infty$, are also expected \cite{Etheredge:2022opl, Etheredge:2023odp, Etheredge:2024tok, Etheredge:2024amg}.
One may then identify DM with a (possibly composite) string heavy state by replacing $\alpha \mapsto - \beta$, with $\beta > 0$.
As an example, we consider both the saturating values $\gamma=2\beta=2/\sqrt{d-2}$.
The stable solution delivers
\begin{equation}
    \epsilon = \dfrac{d-1}{3}.
\end{equation}
Hence, $\epsilon<1$ only for $d=3$.
For $d=4$, the accelerating phase may be arbitrarily long-lived (ignoring the cutoff), approaching $\epsilon = 1$ from below, but the solution would still not feature a cosmic horizon \cite{Marconnet:2022fmx, Andriot:2023wvg, Marconnet:2025vhj}.\footnote{The possibility that EFTs coupled to gravity with a cosmological horizon cannot be UV-complete has been discussed in ref.~\cite{Hassfeld:2025hjx}; see also refs.~\cite{Hebecker:2023qke, Andriot:2023wvg}.}

A large ratio $\beta/\gamma$ favors a small $\epsilon$.
In fact, it is not necessary for the bounds on $\beta$ and $\gamma$ to be saturated (and at the same time).
There is therefore an obvious chance to look for $\epsilon \leq 1$.
For instance, let us consider a 4d compactification on a compact space with constant negative curvature.
A candidate heavy state is a Kaluza-Klein (KK) monopole \cite{Etheredge:2022opl, Rudelius:2022gbz}.
If we assume isotropy, the KK monopole mass and the curvature-induced potential only depend on the Einstein-frame radion, with $\beta = \sqrt{2/3}$ and $\gamma = \sqrt{8/3}$.
In the absence of the DM/DE coupling, the orthogonal direction, i.e. the 10d dilaton $\smash{\tilde{\phi}}$, would be frozen by Hubble friction \cite{Shiu:2023fhb}.
Clearly, this holds even with a DM/DE coupling that is aligned with the potential; see also eq. (\ref{general DMDE epsilon}).
Then, eq.~(\ref{DMDE epsilon}) gives
\begin{equation}
    \epsilon = 1.
\end{equation}
This construction generalizes including further potential terms and letting all moduli be dynamical, but we defer detailed analyses to later studies.
Further candidate states with a growing mass are branes with all their spatial directions wrapped on internal cycles.
These correspond to $d$-dimensional point particles whose mass equals the dimensionally-reduced tension.\footnote{Wrapped branes as DM were considered in ref.~\cite{Shiu:2003ta}.
Their role as heavy states was also explored by ref.~\cite{Casas:2024xqy}.
However, note that in our case we are not studying transient quasi-dS solutions, and we do not explicitly need an EFT realization of DM.}
Backgrounds of cosmic-string loops with increasing tension would also provide a realization of our scenario.\footnote{Loops of fundamental superstrings correspond to the same dynamical system, with $\beta \leq 0$ -- i.e. decreasing tension instead --, as recently studied by ref.~\cite{SanchezGonzalez:2025uco}.}

To conclude, we now illustrate an argument against an epoch of cosmic acceleration that is arbitrarily extended.
With a growing DM mass, a minimal consistency requirement is for the DM Schwarzschild radius to remain smaller than the Hubble radius.
If large black holes were to form out of DM and become comparable in size to the Hubble scale, our construction would cease to be valid; in fact, it would not even be consistent to model DM as a homogeneous fluid well before this point is reached.
Therefore, by imposing the inequality $H^{d-3} m \leq m_{\p,d}^{d-2}$, we find that the black hole size can never outgrow the apparent horizon as long as $2 \beta/\gamma \leq d-3$.
Hence, a necessary condition for large black holes to never form is that $\epsilon \geq 1$.
If we use this criterion as a way to set an upper bound on the duration of the stable solution, we always conclude with a maximum number of Hubble times of accelerated expansion that is possibly large,\footnote{The black-hole/apparent cosmic horizon crossing corresponds to $t = t_0 + \Delta t$, with $\smash{\Delta t/t_0 = (m_{\p,d}^{d-2}/m_0 H_0^{d-3})^{1/[2\beta/\gamma-(d-3)]}- 1}$. Depending on the mass hierarchies, one can find large $\Delta t/t_0$.} but necessarily finite.
This conclusion echoes the arguments supporting the absence of infinitely-extended accelerated-expansion phases based on entropy bounds \cite{Ooguri:2018wrx}.

\section{Asymptotics with DM/DE couplings}

There is a second way to exploit the DM/DE coupling, beyond the one considered so far.
In string compactifications, because of the string-coupling and volume expansions, negative potentials alone may push the EFT outside of perturbative control.\footnote{An interplay between positive- and negative-definite potentials may still yield vacua in regions under perturbative control \cite{Dine:1985he}.
As discussed below, even in constructions with negative-energy vacua (see e.g. ref~\cite{DeWolfe:2005uu}), the DM/DE coupling provides an uplifting potential term of potential cosmological interest.}
A coupling to a light tower of states can however allow for a dynamical evolution towards controlled regions.
In eqs.~(\ref{m}, \ref{V}), let $\smash{\beta = -\alpha < 0}$ and $\Lambda = - K < 0$, still with $\gamma = \gamma_- > 0$ and $\mu > 0$.
Based on eq.~(\ref{effective DMDE-coupled potential}), we find the effective potential
\begin{equation} \label{alternative effective DMDE-coupled potential}
    U_-(\varphi) = - K \, \e^{- \kappa_d \gamma_- \varphi} + n_{\DM, 0} \, \Bigl(\dfrac{a_0}{a}\Bigr)^{d-1} \, \mu \, \e^{- \kappa_d \alpha \varphi}.
\end{equation}
If $\gamma_- > \alpha$, $U_-$ exhibits a runaway towards zero energy at $\varphi = +\infty$; see fig. \ref{fig: complete alternative scenario}.
Instead, $V=V_-$ alone would only have a runaway towards negatively-infinite energy at $\varphi=-\infty$.
The DM particle mass flows to zero, too, remaining in the EFT.
As the scale factor is not fixed, the positive term in eq.~(\ref{alternative effective DMDE-coupled potential}) effectively falls down much more quickly than with the exponential rate $\alpha$.
Even transient epochs can never have $\epsilon \leq 1$, as is obvious from the dynamical system, where now $(y)^2<0$.
If $w=0$, because $(x)^2 + (z)^2 = 1 - (y)^2 > 1$, we have $\epsilon/(d-1) = (x)^2 + (z)^2/2 > 1/2$.

To possibly host epochs of cosmic acceleration, one should also include positive-definite potentials.
One such deformation simply requires an additional potential term $V_+ = \Lambda \, \e^{- \kappa_d \gamma_+ \varphi}$, with $\Lambda > 0$ and $\gamma_+ > \gamma_-$.
Without the DM/DE coupling, the potential would not be a runaway towards $\varphi=+\infty$ and the combination of only the positive potential and the DM/DE-term would just be an exponential runaway.
It is the combination of all three effects that generates a positive-energy runaway towards $\varphi=+\infty$ along with a scale factor-dependent minimum in the effective potential; see fig. \ref{fig: complete alternative scenario}.

\begin{figure}[ht]
    \centering
    
    \begin{tikzpicture}[xscale=1.45,yscale=2.65]

    \draw[->] (-0.85,0) -- (4,0) node[below]{$\varphi$};
    \draw[->] (0,-0.85) -- (0,0.80) node[left]{};

    \draw[domain=0.20:3.5, smooth, thick, variable=\x, orange] plot ({\x}, {-2*exp(-2*\x) + exp(-\x)}) node[above right, black]{};

    \draw[domain=-0.25:3.5, smooth, thick, variable=\x, green!85!blue] plot ({\x}, {-2*exp(-2*\x) + 1.25*exp(-4*\x)});

    \draw[domain=0.65:3.5, smooth, thick, variable=\x, purple!10!cyan] plot ({\x}, {1.25*exp(-4*\x) + exp(-\x)});

    \draw[domain=-0.10:3.5, smooth, thick, variable=\x, purple!50!magenta] plot ({\x}, {-2*exp(-2*\x) + 1.25*exp(-4*\x) + exp(-\x)});
    
    \node at (0.65,-0.25){$U_-$};
    \node[right] at (1.35,-0.20){$V_-+V_+$};
    \node[left] at (-0.15,0.55){$U$};
    \node at (0.35,0.60){$U_+$};

    \end{tikzpicture}
    
    \caption{With the orderings $\gamma_+ > \gamma_- > \alpha > 0$, and at a fixed scale factor, the effective potentials $\smash{U_- = V_- + \rho_\DM}$ (orange)
    and $\smash{U_+ = V_+ + \rho_{\DM}}$ (cyan), the pure field potential $\smash{V_- + V_+}$ (green), and the full effective potential $U$ combining all terms, with a positive-energy minimum (magenta).}
    
    \label{fig: complete alternative scenario}
\end{figure}

\section{Discussion}

A DM/DE coupling can induce a long-lived SEC violation.
The fundamental requirement is for the DM mass and DE potential to depend on the DE field with opposite slopes.
If it exists, the solution is linearly stable, and hence attainable without fine-tuned initial conditions and arbitrarily long-lived in a mathematical sense.
In the scenario we considered, DM may be simply treated as a classical source, circumventing the upper bound set by the EFT cutoff.
Our results may open up new possibilities to overcome the apparent limitations to long-lived SEC violations evidenced in the simplest string compactifications evolving in the field-space asymptotics \cite{Cicoli:2018kdo, Olguin-Trejo:2018zun, Agrawal:2018rcg, Bedroya:2019snp, Hebecker:2019csg, Rudelius:2021azq, Cicoli:2021fsd, Rudelius:2022gbz, Andriot:2022xjh, Calderon-Infante:2022nxb, Shiu:2023nph, Cremonini:2023suw, VanRiet:2023cca}.
In any case, however, the requirement for DM to remain a homogeneous cosmological fluid suggests that the SEC violation cannot last forever.

Our results extend simply to multi-field cosmologies.
Let us consider a multi-dimensional field space, with coordinates $\smash{\varphi = (\varphi^a)}$, with $a=1,\dots,n$, and a multi-exponential positive potential $\smash{V = \sum_i \Lambda_i \, \e^{- \kappa_d \, \gamma_i \cdot \varphi}}$, where each coupling $\gamma_i = (\gamma_{ia})$ is a constant vector in field space, with $i=1,\dots,m$.
In the absence of additional fluids, the attractors are straight lines in field space, with the direction fixed by the minimal-length vector $\gamma_\infty$ joining the origin to the convex hull generated by the  $\gamma_i$-vectors \cite{Shiu:2023fhb}.
In the presence of a coupled DM mass $\smash{m = \mu \, \e^{\kappa_d \, \beta \cdot \varphi}}$, where $\beta = (\beta_a)$ is also a field-space vector, there exists a solution with
\begin{equation} \label{general DMDE epsilon}
    \epsilon = \dfrac{d-1}{2} \dfrac{1+w}{1 + \beta \cdot \gamma_\infty/(\gamma_\infty)^2},
\end{equation}
which is in fact equivalent to projecting the multi-field problem onto $\gamma_\infty$; see fig. \ref{fig.: multi-field DMDE coupling}.
We expect the attractor to be the critical-point solution with the smallest $\epsilon$-value \cite{Shiu:2024sbe}.
By comparing eq.~(\ref{general DMDE epsilon}) with other critical-point solutions, namely the scaling-like solution with $\smash{\epsilon = [(d-2)/4] \, \gamma \cdot \gamma}$, the kination/tracker solution with $\smash{\epsilon = (d-1)(1+w)/2 + (d-2) \, \beta \cdot \beta/(1-w)}$, and the kinating solution with $\epsilon = d-1$ (see app. \ref{app.: multi-field critical points} for details), we may thus predict the late-time solution as the one with the smallest one among such $\epsilon$-values.
For instance, a large-enough value $\smash{\beta \cdot \gamma_\infty / (\gamma_\infty)^2 > 0}$ will make eq.~(\ref{general DMDE epsilon}) the late-time solution.
We also note that our discussion applies to multi-field multi-exponential potentials in spatially-hyperbolic cosmologies: for $w=-(d-3)/(d-1)$ and $\beta=0$, one finds $\epsilon=1$, generalizing the well-known single- \cite{Marconnet:2022fmx, Andriot:2023wvg} and two-field solutions \cite{Marconnet:2025vhj}.

\begin{figure}[ht]
    \centering
    
    \begin{tikzpicture}[scale=1.50,rotate=0]
    
    \draw[orange, thick, fill=orange!18!white] (2,-1.8) -- (2,1.2) -- (2.8,-1.2) -- (2,-1.8);
    
                \draw[->, thick, teal] (0,0) -- (2,-1.8) node[right]{$\gamma_1$};
                \draw[->, thick, teal] (0,0) -- (2,1.2) node[right]{$\gamma_2$};
                \draw[->, thick, teal] (0,0) -- (2.8,-1.2) node[right]{$\gamma_3$};
                \draw[->, thick, purple] (0,0) -- (2,0) node[below left, align=center]{$\gamma_\infty$};
    
                \draw[->, thick, cyan] (0,0) -- (3.2,0.8) node[right]{$\beta$};
                \draw[densely dotted, thin, cyan] (3.2,0.8) -- (3.2,0);
                \draw[->, thin, cyan] (2,0) -- (3.2,0) node[right]{$\beta_\infty = \dfrac{(\beta \cdot \gamma_\infty)}{(\gamma_\infty)^2} \, \gamma_\infty$};

    \end{tikzpicture}
    
    \caption{A representation of the multi-field potential (teal) and DM/DE (blue) couplings, with the asymptotic field direction (purple); one may write $\epsilon = [(d-1)(1+w)/2]/(1+\beta_\infty/\gamma_\infty)$.}
    
    \label{fig.: multi-field DMDE coupling}
\end{figure}

A top-down realization of the scenario we discussed needs to address model-building challenges that deserve further study.
This goes beyond the interests of the current note, focused on the fate of UV-complete long-lived SEC violations.
As an important addition, though, we emphasize that the mechanism we presented works as a way to prolong an epoch of cosmic acceleration even if only a fraction of the total DM has a field-dependent mass.
Any DM fraction with constant mass present at a given time -- including today -- would just eventually redshift away with respect to the DE and field-dependent DM energy densities, reducing the problem to the one discussed in this note.

\begin{acknowledgments}
\subsection*{Acknowledgments}
We are happy to thank Moritz Münchmeyer for a useful discussion on DM/DE-coupled cosmologies.
We are also grateful to Miguel Montero, for testing our proposal with insightful questions on DM as a classical source and black-hole formation from DM, and to Shinji Tsujikawa, for bringing up recent studies of phenomenological implications of DM/DE-coupled cosmologies.
GS is supported in part by the DOE grant DE-SC0017647.
FT is funded by the European Union - NextGenerationEU/PNRR mission 4.1; CUP: C93C24004950006.
HVT is supported in part by NSF grant DMS-2348305.
\end{acknowledgments}

\appendix

\section{Linear stability} \label{app.: linear stability}

Here, we discuss the linear stability of the DM/DE-coupled solution.
Specifically, we get rid of the function $h$ and its variation via the time reparameterization $\smash{\sigma - \sigma_0 = \int_{t_0}^t \de t' \, h(t')}$ and of the variable $y$, ignoring its variation, too, after implementing the topological constraint into the remaining equations.
We thus have the autonomous system of ordinary differential equations
\begin{subequations}
\begin{align}
    \dfrac{\de x}{\de \sigma} & = - (x - c) [1-(x)^2-Z] - \dfrac{1-w}{2} \, x Z - b Z, \label{reduced DMDE x-equation} \\
    \dfrac{\de Z}{\de \sigma} & = 2 \biggl[ - \dfrac{1+w}{2} + (x)^2 + \dfrac{1+w}{2} \, Z + b x \biggr] Z, \label{reduced DMDE z2-equation}
\end{align}
\end{subequations}
with $Z = (z)^2$.
It is assumed that $w \in\; ]1,1[$ and, without loss of generality, we fix $c>0$.
We focus on the critical points discussed in the main text, i.e.
\begin{subequations}
\begin{align}
    \mathrm{x} & = \dfrac{1}{2} \, \dfrac{1+w}{b + c}, \label{reduced xCP} \\
    \mathrm{Z} & = \dfrac{2c(b+c) - (1+w)}{2 (b+c)^2}, \label{reduced z2CP}
\end{align}
\end{subequations}
with the necessary existence conditions
\begin{subequations}
\begin{align}
    \chi_1 & = c (b + c) - \dfrac{1+w}{2} > 0, \label{reduced DMDE CP existence 1} \\
    \chi_2 & = b (b + c) + \dfrac{1 - w^2}{4} > 0. \label{reduced DMDE CP existence 2}
\end{align}
\end{subequations}
One may also observe that $\chi_0 = b+c > 0$ is implied by eq.~(\ref{reduced DMDE CP existence 1}).
By linearizing the dynamical system around the critical point, after defining
\begin{equation}
    \zeta = b + \dfrac{1-w}{2} \, c,
\end{equation}
we see that the stability matrix has eigenvalues
\begin{equation}
    \lambda_\pm = \dfrac{- \zeta \pm \sqrt{\zeta^2 - 8 \chi_1 \chi_2}}{2 \chi_0}.
\end{equation}
By the existence conditions in eqs.~(\ref{reduced DMDE CP existence 1}, \ref{reduced DMDE CP existence 2}), we can see that $\zeta > 0$; moreover, we know that $\chi_1 \chi_2 > 0$ and $\chi_0 > 0$.
Hence, we see that $\smash{\mathrm{Re} \, \lambda_\pm < 0}$, which means that the critical point is linearly stable.

Besides the critical points discussed above, there are more, corresponding to the solutions where $\mathrm{y}$ and/or $\mathrm{z}$ are set to vanish.

\begin{itemize}[wide, labelwidth=!, labelindent=0pt, parsep=3pt, itemsep=0pt,leftmargin=12pt]
    \item There is a scaling-like critical point
    \begin{subequations}
        \begin{align}
            \mathrm{x} & = c, \\
            \mathrm{Z} & = 0,
        \end{align}
    \end{subequations}
    whose conditions for linear stability are $\smash{c^2 - 1}$ and $\smash{c (c+b) < (1+w)/2}$.
    These correspond to $\smash{\xi = c^2}$.
    \item There is a kination/tracker critical point such that $\smash{(\mathrm{x})^2 + \mathrm{Z}=1}$, with
    \begin{equation}
        \mathrm{x} = - \dfrac{2b}{1-w}.
    \end{equation}
    This is linearly-stable for $\smash{b (b+c) < - (1-w^2)/4}$ and $\smash{b^2 < (1-w)^2/4}$, and it corresponds to $\smash{\xi = (1+w)/2 + 2 b^2/(1-w)}$.
    \item There is a kinating solution
    \begin{equation}
        \mathrm{x} = 1,
    \end{equation}
    whose conditions for linear stability are $\smash{c>1}$ and $\smash{b < - (1-w)/2}$, with $\xi = 1$.
    There is a second kinating solution, i.e. $\smash{\mathrm{x}=-1}$, that however is always unstable (linear stability would require $\smash{1+c<0}$ and $\smash{b > (1-w)/2}$).
\end{itemize}

A complete representation of the linearly-stable solutions in parameter space, including the corresponding $\epsilon$-parameter, is in fig. \ref{fig.: critical-point classification}.

\begin{figure}[H]
    \centering
    
    \begin{tikzpicture}[scale=2.5,every node/.style={font=\normalsize}]

    \fill[teal,opacity=0.12] (-0.5,1) rectangle (-1.4,1.7225);

    \draw[name path = C1, domain=-0.5:1.45, smooth, thick, variable=\b, , magenta, thick, opacity=0] plot ({\b}, {(-\b+sqrt((\b)^2+2))/2});
    \draw[name path = ckin, domain=-1.4:-0.5, smooth, thick, variable=\b, magenta, thick, opacity=0] plot ({\b}, {1});
    \draw[name path = czero1, domain=-0.5:1.45, smooth, thick, variable=\b, orange, opacity=0] plot ({\b}, {0});
    \draw[name path = czero2, domain=-1.4:-0.5, smooth, thick, variable=\b, orange, opacity=0] plot ({\b}, {0});
    \tikzfillbetween[of = C1 and czero1]{orange,opacity=0.12};
    \tikzfillbetween[of = ckin and czero2]{orange,opacity=0.12};

    \draw[name path = C2, domain=-0.5:-0.16, smooth, thick, variable=\b, magenta, thick, opacity=0] plot ({\b}, {-0.25/\b-\b});
    \draw[name path = t, domain=-0.5:-0.16, smooth, thick, variable=\b, teal, opacity=0] plot ({\b}, {1.7225});
    \tikzfillbetween[of = C2 and t]{yellow!85!red!95!black,opacity=0.35};

    \tikzfillbetween[of = C1 and C2]{green!50!teal,opacity=0.12};
    \fill[color=green!50!teal, opacity=0.12] (-0.16,1.7225) -- (1.45,1.7225) -- (1.45,0.2877) -- (-0.16,1.7225);

    \draw[thick,magenta] (-0.5,1) -- (-0.5,1.7225);
    \draw[domain=-0.5:1.45, smooth, thick, variable=\b, magenta, thick, opacity=1] plot ({\b}, {(-\b+sqrt((\b)^2+2))/2}); 
    \draw[domain=-1.4:-0.5, smooth, thick, variable=\b, magenta, thick, opacity=1] plot ({\b}, {1}); 
    \draw[domain=-0.5:-0.16, smooth, thick, variable=\b, magenta, thick, opacity=1] plot ({\b}, {-0.25/\b-\b}); 

    \node[teal, align=center] at (-0.95,1.35) {$\epsilon = d-1$};
    \node[orange, align=center] at (-0.95,0.40) {$\epsilon = \dfrac{d-2}{4} \gamma^2$};
    \node[yellow!85!red!95!black] at (-0.25,1.90) {$\epsilon = \dfrac{d-1}{2} \, \Bigl[ 1+w + \dfrac{d-2}{d-1} \dfrac{\beta^2}{1-w} \Bigr]$};
    \node[green!50!teal, align=center] at (0.725,0.92) {$\epsilon = \dfrac{d-1}{2} \dfrac{1+w}{1 + \beta/\gamma}$};
        
    \draw[dotted] (-0.5,1) -- (0,1) node[above left] {$\Gamma_d$};
    \draw[dotted] (-0.5,1) -- (-0.5,0) node[below] {$-\dfrac{1-w}{2} \Gamma_d$};

    \draw[->] (-1.55,0) -- (1.60,0) node[below]{$\beta$};
    \draw[->] (0,-0.16) -- (0,1.55) node[left]{$\gamma$};
    \node[below left] at (0,0) {$0$};

\end{tikzpicture}
    
\caption{The classification of the linearly-stable critical points in parameter space: the DM/DE-coupled solution (green), the scaling-like solution (orange), the kination/tracker solution (yellow), and a kinating solution (teal).}
    
\label{fig.: critical-point classification}
\end{figure}

\section{Multi-field critical points} \label{app.: multi-field critical points}

Let us consider a multi-field theory with couplings
\begin{subequations}
    \begin{align}
        m & = \mu \, \e^{\kappa_d \beta_a \varphi^a}, \label{multi-field m} \\
        V & = \sum_i \Lambda_i \, \e^{- \kappa_d \gamma_{ia} \varphi^a}. \label{multi-field V}
    \end{align}
\end{subequations}
The cosmological equations read
\begin{subequations}
\begin{align}
    & \ddot{\varphi}^a + (d-1) H \dot{\varphi}^a + \dfrac{\der V}{\der \phi_a} = - \dfrac{\rho}{m} \dfrac{\der m}{\der \phi_a}, \label{multi-field DMDE FRW-KG eq.} \\
    & \dot{\rho} + (d-1) (1+w) H \rho = \dfrac{\rho}{m} \, \dfrac{\der m}{\der \varphi^a} \, \dot{\varphi}^a, \label{multi-field DMDE continuity eq.} \\
        & H^2 = \dfrac{2 \kappa_d^2}{(d-1) (d-2)} \biggl[ \dfrac{1}{2} \, \dot{\varphi}_a \dot{\varphi}^a + V + \rho \biggr], \label{multi-field DMDE Friedmann eq.}
\end{align}
\end{subequations}
with $\smash{\dot{H} = - [2 \kappa_d^2/(d-2)] \, \bigl[ \dot{\varphi}_a \dot{\varphi}^a/2 + [(1+w)/2] \, \rho \bigr]}$, too.

We define $h = (d-1) H$, $b_a = \beta_a/\Gamma_d$ and $c_{ia} = \gamma_{ia}/\Gamma_d$, with $\smash{\Gamma_d = 2 \sqrt{d-1}/\sqrt{d-2}}$, and
\begin{align*}
    x^a & = \dfrac{\kappa_d}{\sqrt{d-1} \sqrt{d-2}} \, \dfrac{\dot{\varphi}^a}{H}, \\
    y_i & = \dfrac{\kappa_d \sqrt{2}}{\sqrt{d-1} \sqrt{d-2}} \, \dfrac{1}{H} \, \sqrt{\Lambda_i \, \e^{- \kappa_d \gamma_{ia} \varphi}}, \\
    z & = \dfrac{\kappa_d \sqrt{2}}{\sqrt{d-1} \sqrt{d-2}} \, \dfrac{\sqrt{\rho}}{H}.
\end{align*}
Then, eqs.~(\ref{multi-field DMDE FRW-KG eq.}, \ref{multi-field DMDE continuity eq.}, \ref{multi-field DMDE Friedmann eq.}) become
\begin{subequations}
\begin{align}
    \dot{x}_a & = f(x,y,z) \, h, \label{multi-field DMDE x-equation} \\[1.8ex]
    \dot{y}^i & = \biggl[ (x)^2 + \dfrac{1+w}{2} \, (z)^2 - c_{ia} x^a \biggr] \, y h, \label{multi-field DMDE y-equation} \\[0.45ex]
    \dot{z} & = \biggl[ - \dfrac{1+w}{2} + (x)^2 + \dfrac{1+w}{2} \, (z)^2 + b_a x^a \biggr] \, z h, \label{multi-field DMDE z-equation} \\
    \dot{h} & = - \biggl[ (x)^2 + \dfrac{1+w}{2} \, (z)^2 \biggr] h^2, \label{multi-field DMDE h-equation}
\end{align}
\end{subequations}
where for brevity we defined
\begin{align*}
    f(x,y,z) = - \sum_{i} (x_a - c_{ia}) (y^i)^2 - \Bigl(\dfrac{1-w}{2} \, x_a + b_a \Bigr) (z)^2,
\end{align*}
jointly with the condition
\begin{equation}
    (x)^2 + (y)^2 + (z)^2 = 1. \label{multi-field DMDE sphere condition}
\end{equation}

Besides those where $\mathrm{y}_i=0$ and/or $\mathrm{z}=0$, the search of critical points is algebraically demanding.
In particular, we restrict our attention to cases where the matrix $c_{ia}$ is square, knowing that, for higher-dimensional matrices, one simply has to define the square matrix of the couplings that define the shortest coupling convex-hull distance \cite{Shiu:2023fhb, Shiu:2024sbe}.
It is convenient to define
\begin{subequations}
    \begin{align}
        k & = \sum_{i} \sum_{j} [(c^{-1})^{\mathrm{t}}]^i{}_a (c^{-1})^{a j}, \\
        c_{\infty}^a & = \dfrac{1}{k} \, \sum_{i} (c^{-1})^{a i},
    \end{align}
\end{subequations}
where we also notice the identity $\smash{c_{\infty a} c_{\infty}^a = 1/k}$.
Then, the novel critical points read
\begin{subequations}
    \begin{align}
        \mathrm{x}^a & = \dfrac{1+w}{2} \, \dfrac{k (b_a c_{\infty}^a)}{1 + k (b_a c_{\infty}^a)}, \label{multi-field xCP} \\
        (\mathrm{z})^2 & = \dfrac{1 + k (b_a c_{\infty}^a) - \dfrac{1+w}{2} \, k}{[1 + k (b_a c_{\infty}^a)]^2}, \label{multi-field zCP}
    \end{align}
\end{subequations}
which correspond to the $\xi$-parameter
\begin{equation} \label{multi-field xiCP}
    \xi = \dfrac{1+w}{2} \, \dfrac{1}{1 + k (b_a c_{\infty}^a)}.
\end{equation}
This corresponds to the $\epsilon$-parameter in the multi-field discussion in the main text, along with its geometric interpretation.
The additional solutions are as follows.
For $\mathrm{z}=0$, we find the usual scaling-like critical point \cite{Collinucci:2004iw, Shiu:2023fhb}, with $\xi=1/k$.
For $\mathrm{y}_i=0$, we find the multi-field kination/tracker solution with $\xi = (1+w)/2 + 2 (b_a b^a)/(1-w)$.
For $\mathrm{z}=0$ and $\mathrm{y}_i=0$, we find the multi-field kinating solution with $\xi=1$.

\section{Wrapped branes as \texorpdfstring{\\}{} string-theoretic heavy states} \label{app.: wrapped branes as string-theoretic heavy states}

Obvious candidates for states with a growing mass are branes with all their spatial directions wrapped on internal cycles.
These appear as $d$-dimensional point particles whose mass equals the dimensionally-reduced tension.

As an illustration, we consider an anisotropic compactification on two $s_{1,2}$-dimensional spaces, with $s_1+s_2=10-d$.
In terms of canonical fields, the metric is
\begin{equation}
    ds_{1,9}^2 = \e^{\frac{2 \kappa_d \tdelta}{\sqrt{d-2}}} d \hat{s}_{1,d-1}^2 + \e^{\frac{2 \kappa_d \tsigma_1}{\sqrt{s_1}}} d \breve{s}_{s_1}^2 + \e^{\frac{2 \kappa_d \tsigma_2}{\sqrt{s_2}}} d \breve{s}_{s_2}^2,
\end{equation}
where we assume the line elements $\smash{d \hat{s}_{1,d-1}^2}$ and $\smash{d \breve{s}_{s_{1,2}}^2}$ to be fixed.
For a D$p$-brane wrapping $n_{1,2}$-directions in the $s_{1,2}$-spaces, with $n_1+n_2=p$, or, in an analogous fashion, an NS5-brane, the particle masses read
\begin{subequations}
    \begin{align}
        m_{\mathrm{D}p} & = \mu_{\mathrm{R}} \, \e^{-\frac{d-4}{2 \sqrt{d-2}} \kappa_d \tdelta + \frac{2 n_1 - s_1}{2 \sqrt{s_1}} \kappa_d \tsigma_1 + \frac{2 n_2 - s_2}{2 \sqrt{s_2}} \kappa_d \tsigma_2}, \\
        m_{\mathrm{NS}5} & = \mu_{\mathrm{NS}} \, \e^{-\frac{d-3}{\sqrt{d-2}} \kappa_d \tdelta + \frac{n_1 - s_1}{\sqrt{s_1}} \kappa_d \tsigma_1 + \frac{n_2 - s_2}{\sqrt{s_2}} \kappa_d \tsigma_2},
    \end{align}
\end{subequations}
for reference values $\smash{\mu_{\mathrm{R}}}$ and $\smash{\mu_{\mathrm{NS}}}$.

For instance, if the $s_2$-space has constant negative curvature, a positive-definite potential
\begin{align*}
    V_{R_2} = \Lambda_{R_2} \, \e^{\frac{2 \kappa_d \tdelta}{\sqrt{d-2}} - \frac{2 \kappa_d \tsigma_2}{\sqrt{s_2}}} = \Lambda_{R_2} \, \e^{- \frac{2 \sqrt{8 - s_1}}{\sqrt{s_2} \sqrt{d-2}} \kappa_d \varphi}
\end{align*}
is generated, where an $\mathrm{O}(2)$-transformation singles out a unique canonical scalar $\varphi$ the potential depends on, with an orthogonal flat direction $\xi$.
Let us assume that the moduli $\smash{\tsigma_1}$ and $\xi$ are frozen by Hubble friction.
Then, we can project the DM masses on the $\varphi$-direction, which curiously both have the dependence
\begin{align*}
    m_{\mathrm{D}p, \mathrm{NS}5} & = \mu_{\mathrm{R}, \mathrm{NS}} \, \e^{\frac{(d-2) n_2 - s_2}{\sqrt{d-2} \sqrt{s_2} \sqrt{8 - s_1}} \, \kappa_d \varphi}.
\end{align*}
If $n_2 \geq s_2/(d-2)$, we find $\beta, \gamma > 0$,
with
\begin{align*}
    \epsilon = \dfrac{(d-1) (8-s_1)}{(d-2) (n_2+2) + s_2}.
\end{align*}
One can verify that the inequality $\epsilon \leq 1$ can never be fulfilled in this example.
Whether constructions of this kind -- and more complex versions thereof -- can be embedded in a full string realization with $\epsilon \leq 1$ requires further work.
For now, we defer a systematic analysis.

\bibliographystyle{apsrev4-1}
\bibliography{refs.bib}

\end{document}